\documentclass[aps,pra,reprint]{revtex4-1}

\usepackage{amsmath}
\usepackage{amssymb}
\usepackage{graphicx}
\usepackage{hyperref}
\usepackage{blindtext}

\begin{abstract}
Tightly-focused, ultrashort radially-polarized laser beams have a large longitudinal field, which provides a strong motivation for direct particle acceleration and manipulation in vacuum. The broadband nature of these beams means that chromatic properties of propagation and focusing are important to consider. We show via single particle simulations that using the correct frequency-dependent beam parameters is imperative, especially as the pulse duration decreases to the few-cycle regime. The results with different spatio-spectral amplitude profiles show both a drastic increase or decrease of the final accelerated electron energy depending on the shape, motivating both proper characterization and potentially a route to optimization.
\end{abstract}
\begin{document}

\title{On the importance of frequency-dependent beam parameters for vacuum acceleration with few-cycle radially-polarized laser beams}
\author{Spencer W. Jolly}
\email{spencer.jolly@cea.fr}
\affiliation{LIDYL, CEA, CNRS, Universit{\'e} Paris-Saclay, CEA Saclay, 91 191 Gif-sur-Yvette, France}

\date{\today}

\maketitle

The longitudinal acceleration of electrons with high-power tightly-focused radially-polarized laser beams (RPLBs) is a promising method for particle acceleration in vacuum and provides a rich platform for studying relativistic particle dynamics and the proper modeling of the focused fields~\cite{varin05,fortin10,wong10}. Simulations to date have progressed to use representations for the field that are accurate non-paraxially to high order~\cite{salamin06,marceau12,marceau13-1,marceau13-2,martens14} and have even included inter-particle interactions and radiation reaction models~\cite{wong17}. Experimental results have been impressive~\cite{payeur12,carbajo16}, but still require an enhanced knowledge of the complete field to understand the limitations and to predict future results. Recent attempts have been made to show the effect of chromatic focusing and linear chirp~\cite{jolly19-1}, and the potential of a large number of other aberrations are so far unexplored. However, these simulations have all so far not considered the amplitude and phase effects that become important in the diffraction of very short pulses approaching the few-cycle regime~\cite{porras98,feng98,caron99,porras02}. This manuscript will simulate the effect of the frequency-dependent beam parameters on acceleration with few-cycle RPLBs, showing a significant effect. This has been explored for a single case of sub-cycle and single-cycle pulses~\cite{cai16,cai18}, where we show rather the effect for various designed spatio-spectral profiles at many laser powers and pulse durations.

\begin{figure}[b]
	\centering
	\includegraphics[width=84mm]{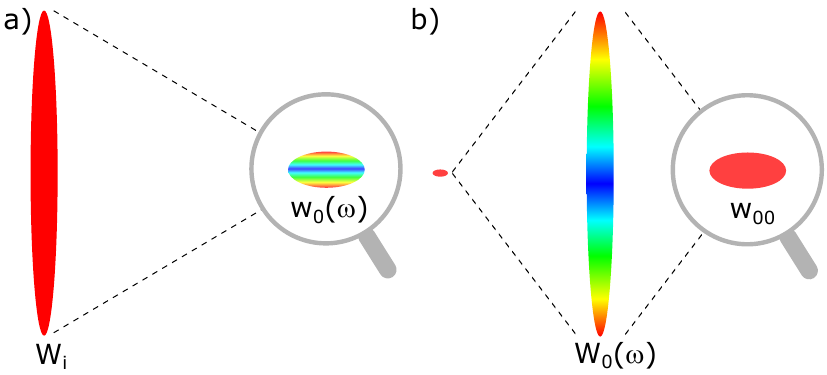}
	\caption{Two examples of unique physical scenarios that require different descriptions in the focus in order to properly model the field. In (a) the collimated beam has a constant beam size, but in-focus the beam size depends on frequency. In (b) it is reversed.}
	\label{fig:fig1}
\end{figure}

Diffraction effects become significant in the few-cycle regime for a very simple reason, because diffraction---the crucial mechanism to model when tightly focusing any laser beam---is itself chromatic. When a laser field is very broadband the chromatic nature of diffraction becomes relevant, affecting both the pulse amplitude and phase through the focus~\cite{porras02}. The effect has been measured on-axis and off-axis for linearly polarized pulses using a variety of techniques~\cite{lindner04,tritschler05,major15,hoff17}, and can be conceptualized by comparing two distinct physical scenarios of laser beam production and focusing, shown in Fig.~\ref{fig:fig1}.

If a laser beam is synthesized in such a way that on the collimated beam all frequencies have the same beam waist $W_i$ and a Gaussian spatial profile, when focused with a perfect focusing element having focal length $f$ each frequency will have a beam size $w_0(\omega)=\frac{2cf}{\omega W_i}$ in the focus (see Fig.~\ref{fig:fig1}(a)). This matters because the in-focus field is generally the relevant field for simulations or experiments. In contrast, if a laser beam is synthesized in way such that all frequencies have the same waist $w_{00}$, and then collected, the collimated beam will have the waist $W_0(\omega)=\frac{2cf}{\omega w_{00}}$. In this second case when the large collimated beam is re-focused, the waist will be constant again in the experimental focus (see Fig.~\ref{fig:fig1}(b)).

As an even further complication, the Rayleigh range is not constant in either of these two cases since $z_R=\frac{\omega w_{0}^2(\omega)}{2c}$ depends on frequency. This underscores the lack of validity of the common model of pulse focusing, which usually operates on the approximation that both the beam waist and Rayleigh range are frequency-independent quantities related to the central frequency. As the pulse duration decreases (bandwidth increases) the common approximation becomes less accurate. In this work we show that in vacuum acceleration simulations with few-cycle RPLBs it is crucial to consider the correct model of frequency-dependence of the beam parameters corresponding to the physical scenario in question, whatever it may be.

The relationships between the beam waist and Rayleigh range on the input beam, $W_0$ and $Z_R$ respectively, and on the focused beam, $w_0$ and $z_R$ respectively, are as follows: $Z_R=\frac{\omega W_0^2}{2c}$, $z_R=\frac{\omega w_0^2}{2c}$, and $w_0=\frac{2cf}{\omega W_0}$. Any one parameter of {$W_0$, $Z_R$, $w_0$, $z_R$} may depend arbitrarily on frequency, but the relationships are fixed by diffraction theory. This is crucial for understanding the various situations presented throughout the rest of this manuscript.

The two physical scenarios shown in Fig.~\ref{fig:fig1} are differentiated by the frequency-dependent beam waist of collimated input beam. But in fact, such situations can be classified by a single parameter $g_0$ referred to as the "Porras factor"~\cite{porras02,hoff17}

\begin{equation}
g_0 = \frac{dZ_R(\omega)}{d\omega}\biggr\rvert_{\omega_0} \frac{\omega_0}{Z_R(\omega_0)} .
\end{equation}

\noindent With this new factor in mind we can refer to the different scenarios more easily. The case where the input beam size $W_0$ is constant corresponds to $g_0=1$, which also is a case where the beam is iso-diffracting in the focus (i.e. $w_0/z_R$ is constant), and is the same as Fig.~\ref{fig:fig1}(a). The case where the focused beam size $w_0$ is constant corresponds to $g_0=-1$, where the input beam is iso-diffracting (i.e. $W_0/Z_R$ is constant), and is the same as Fig.~\ref{fig:fig1}(b). A third case $g_0=0$ is where the Rayleigh ranges both on the input beam and the focused beam are frequency-independent ($Z_R=f^2/z_{R0}$). Keep in mind that none of these cases correspond to the simplification that both the beam waist and Rayleigh range are frequency-independent.

More generally, if we define $w_0(\omega_0)=w_{00}$ as the reference value, since we often model the interaction in the focal region and the beam size is easiest to measure, this results in $z_R(\omega)=(\omega_0 w_{00}^2/2c)(\omega_0/\omega)^{g_0}$ and $w_0(\omega)=w_{00}(\omega_0/\omega)^{\frac{g_0+1}{2}}$. Complete descriptions of the phase at arbitrary longitudinal and transverse locations throughout the focal volume have been derived for linearly polarized light~\cite{porras02,porras09,porras18}, but we will consider rather the longitudinal field produced from focusing radially-polarized Gaussian light fields for the acceleration application.

We will first study only on-axis acceleration with RPLBs, where the electron is constrained to a trajectory only on $r=0$ where the purely $E_z$ field is responsible for acceleration. Then we will look at off-axis acceleration as well, where the fully frequency-dependent model becomes even more important. For all simulations we will use a focused beam waist defined at the central frequency as $w_0(\omega_0)=w_{00}=4$\,$\mu$m and calculate the Rayleigh range and beam waist at different frequencies corresponding to the $g_0$ value under consideration. All simulations will be done with pulses having a central wavelength $\lambda_0$ of 800\,nm ($\omega_0=2\pi\lambda_0/c$).

The on-axis longitudinal electric field in frequency space is modelled within the paraxial approximation according to past work~\cite{jolly19-1}

\begin{align}
\hat{E}_z(z,\omega)&=\sqrt{\frac{8P_0}{\pi\epsilon_0{c}}}\frac{A(\omega)}{z_R(\omega)\left(1+\left(\frac{z}{z_R(\omega)}\right)^2\right)}e^{i\psi(z,\omega)} \label{eq:field}\\
\psi(z,\omega)&=\Psi_0+2\tan^{-1}\left(\frac{z}{z_R(\omega)}\right)-\frac{\omega{z}}{c} \label{eq:phase},
\end{align}

\noindent where $A(\omega)$ is the integrated frequency content of the pulse, $z_R(\omega)$ is the in-focus Rayleigh range, $P_0$ is the beam power, $\Psi_0$ is the constant CEP offset phase, and $\epsilon_0$ and ${c}$ are physical constants. The modeling in frequency space is necessary because when $z_R$ depends on frequency, which we have shown to be necessary to properly describe commonly existing scenarios, the field in time is not simple to describe and will have a modified amplitude and phase compared to the frequency-independent case. The paraxial approximation which produces Eqs.~(\ref{eq:field})--(\ref{eq:phase}) is only valid for the on-axis acceleration that we present first~\cite{marceau13-1}.

The electric field in time, calculated via the Fourier transform, is necessary to simulate the electron acceleration via the relativistic Lorentz force and a 5th-order Adams-Bashforth finite-difference method as in much past work~\cite{wong10,jolly19-1}. The simulations start at a large enough negative time such that the pulse is not influencing the electron and they continue until the pulse has completely overtaken the particle and the kinetic energy is no longer changing significantly. We study the final kinetic energy of electrons initially at rest for a range of laser powers, pulse durations, and $g_0$ values, optimized over the initial position $z(0)$ of the electron and the CEP phase $\Psi_0$ of the laser pulse.

\begin{figure}[t]
	\centering
	\includegraphics[width=84mm]{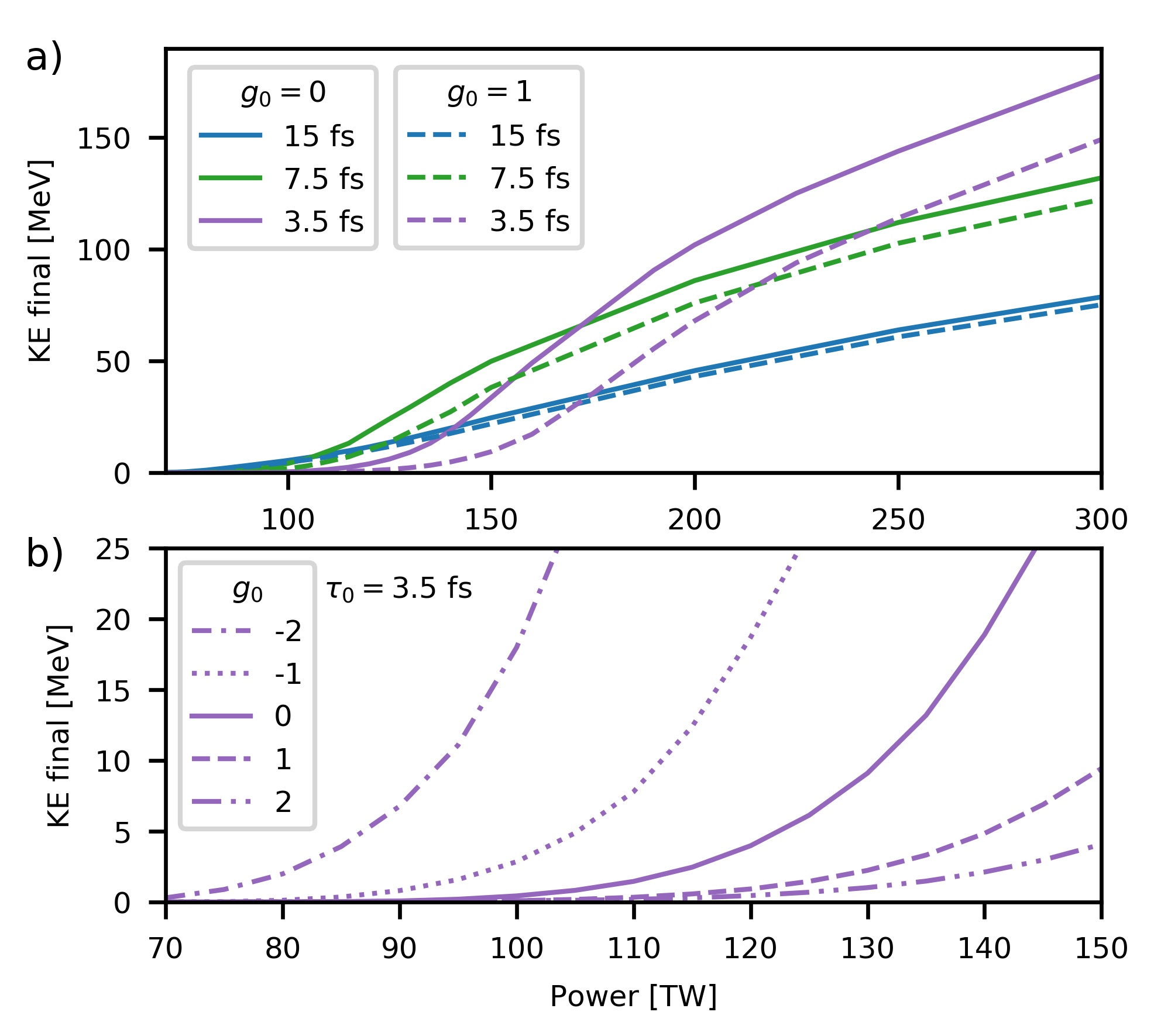}
	\caption{(a) Summary of on-axis results at durations of 15, 7.5, and 3.5\,fs with the constant $z_{R0}$ ($g_0=0$) assumption (solid lines) and with $g_0=1$ (dashed lines). (b) On-axis results at a duration of 3.5\,fs for a variety of $g_0$ values.}
	\label{fig:fig2}
\end{figure}

We first look at laser pulses that have a Gaussian spectrum with characteristic $1/e^2$ spectral width $\Delta\omega$ and pulse duration $\tau_0=2/\Delta\omega$. Using these Gaussian pulses we study acceleration with laser pulses having durations of 15 and 7.5\,fs up to powers of 300\,TW, comparing the default scenario of a constant $z_{R0}$ ($g_0=0$) to that with $g_0=1$. We choose $g_0=1$ since it correponds to the scenario in Fig.~\ref{fig:fig1}(a), which we consider to be a very likely scenario for lasers having these power levels. We extend this analysis to laser pulses of 3.5\,fs duration that have a Poisson-like spectrum, which is necessary to properly model the pulse according to Maxwell's equations as the pulse approaches the few-cycle regime~\cite{caron99}. The results are shown in Fig.~\ref{fig:fig2}(a).

The key conclusions of the results presented in Fig.~\ref{fig:fig2}(a) are that the $g_0=1$ scenario decreases the final kinetic energy in all cases relative to the $g_0=0$ case, and that the relative decrease becomes more severe as the pulse duration decreases. At 130\,TW power in the case of 3.5\,fs duration the decrease is by 75\,\%, where with 15\,fs duration the decrease is only by 13\,\%. For all durations this decrease becomes less severe as the laser power increases. At 300\,TW the decrease is 16\,\% and 4.5\,\% for 3.5\,fs and 15\,fs respectively.

We also consider acceleration with beams having a larger set of $g_0$ values for the shortest pulse duration of 3.5\,fs, with the results shown in Fig.~\ref{fig:fig2}(b). We include the results already presented at $g_0=0$ and 1 for this duration, and expand to include -2, -1, and 2 in the range of lasers powers from 70--150\,TW. The surprising results are that, where the $g_0=1$ case produed a lower final kinetic energy, the cases of negative $g_0$ produce a higher final kinetic energy, increasing as $g_0$ becomes more negative. When $g_0=2$ the final kinetic energy is massively decreased.

\begin{figure}[t]
	\centering
	\includegraphics[width=84mm]{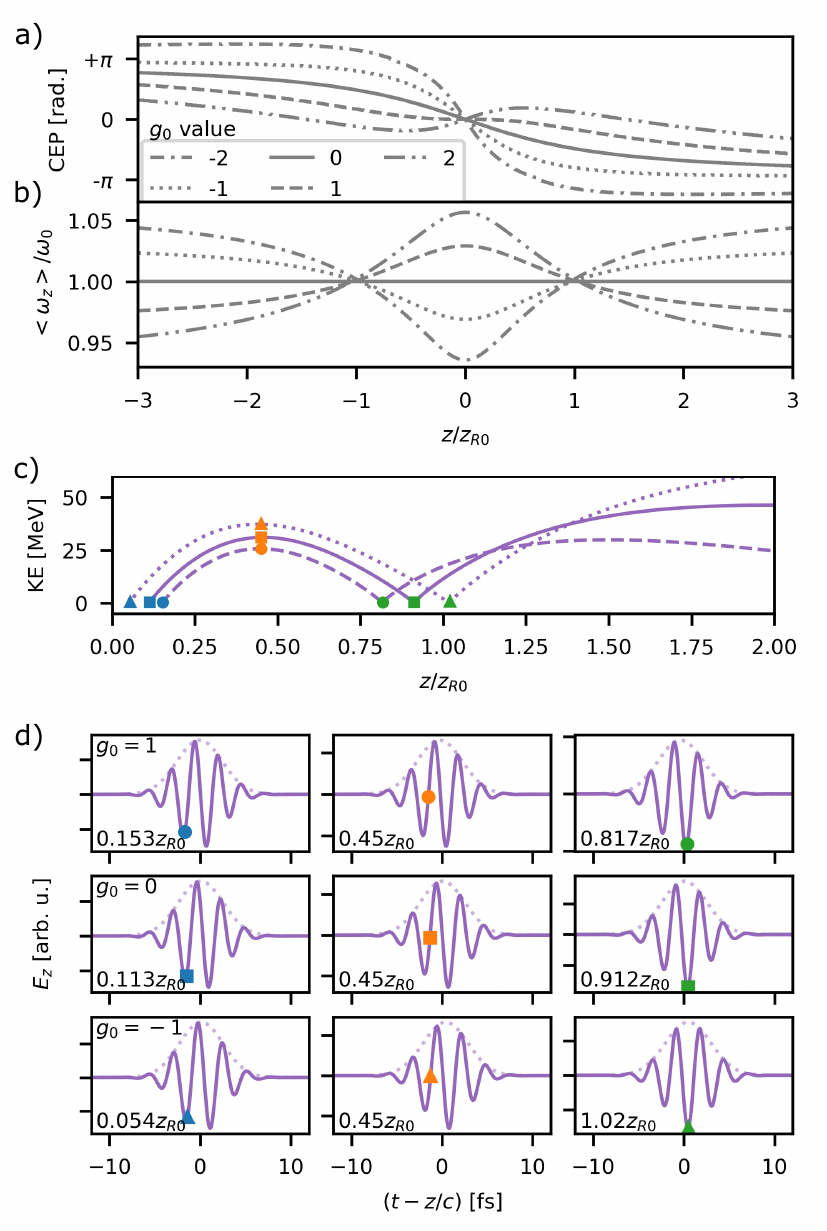}
	\caption{The Gouy phase behavior (a) and the central frequency $\omega_z$ (b) of $E_z$ on-axis depend strongly on the $g_0$ value. Three trajectories at 125\,TW (c) show the different nature of the acceleration (each case having a different optimum $z(0)$). Snapshots of the field (d), with the location of the electron shown with the symbols (circles, squares, triangles), in each case having a different optimum constant $\Psi_0$, show the case of $g_0=-1$ having the best CEP at the start of the final accelerating half-cycle, resulting in more acceleration. The positions of the snapshots in (d) are shown in (c) with the corresponding symbols and colors.}
	\label{fig:field}
\end{figure}

The explanation for this drastic difference in the final kinetic energy must be related to the change in the Gouy phase and central frequency within the focal area for different $g_0$ values, as derived in Ref.~\cite{porras09} and measured in Ref.~\cite{hoff17} for linearly-polarized pulses. We numerically calculate the CEP and central frequency change through the focus for the on-axis $E_z$ field of the focused radially-polarized pulses, with the results shown in Fig.~\ref{fig:field}(a)--(b). The Gouy phase results notably show a similar behavior as in the previous work with linear polarization, i.e. increasing steepness as $g_0$ becomes negative and an inflection as $g_0$ become positive, but with a total shift of $2\pi$. This $2\pi$ shift conflicts with previous work on the Gouy phase in highly non-paraxial focusing~\cite{pang13,kaltenecker16,woldegeorgis18}, where they observed only a shift of $\pi$, but since we are in the paraxial regime this is to be expected.

The impact of the combination of the slightly modified spatio-spectral amplitude and the Gouy phase is that the field at a given time and position is different for the various $g_0$ values, impacting the acceleration as the electron slips behind the laser pulse. We demonstrate this by looking at three acceleration trajectories for $g_0$ values of 1, 0, and -1 at 125\,TW shown in Fig.~\ref{fig:field}(c). Each case has its own optimum constant offset phase $\Psi_0$ and initial electron position $z(0)$. Taking snapshots of the field at the points in each case where the accelerating field is strongest in the second-to-last accelerating half-cycle (left collumn of Fig.~\ref{fig:field}(d)), where the field switches from accelerating to decelerating (middle collumn of Fig.~\ref{fig:field}(d)), and where the field is strongest in the last accelerating half-cycle (right collumn of Fig.~\ref{fig:field}(d)) explains the difference in acceleration. The CEP for the case of $g_0=-1$ in the lower row of Fig.~\ref{fig:field}(d) is such that the decelerating field is below the maximum that it could be when the electron enters the decelerating half-cycle, and when the electron enters the final accelerating half-cycle the CEP has evolved such that the field is at it's maximum possible. The opposite is for the case of $g_0=1$ in the upper row of Fig.~\ref{fig:field}(d), resulting in less acceleraton. The evidence in Fig.~\ref{fig:field}(d) shows that the different CEP evolution shown in Fig.~\ref{fig:field}(a) results in a more favorable situation when $g_0$ is negative.

The analysis of on-axis accleration was instructive and clearly showed that as the pulse duration decreases, proper modeling becomes crucial. The same discussion of including the frequency-dependence of the beam parameters is important for off-axis simulations as well, but there is another dimension that becomes relevant. Where the on-axis field in Eqs.~(\ref{eq:field})--(\ref{eq:phase}) only included the Rayleigh range, the off-axis field equations include independently the beam waist and the Rayleigh range~\cite{salamin06}. Therefore the off-axis field equations in the frequency-independent approximation are lacking the proper form for either the beam waist or the Rayleigh range (since they cannot both be frequency-independent, as discussed earlier), and therefore do not correspond exactly to any physical scenario regardless of the $g_0$ value.

\begin{figure}[t]
	\centering
	\includegraphics[width=84mm]{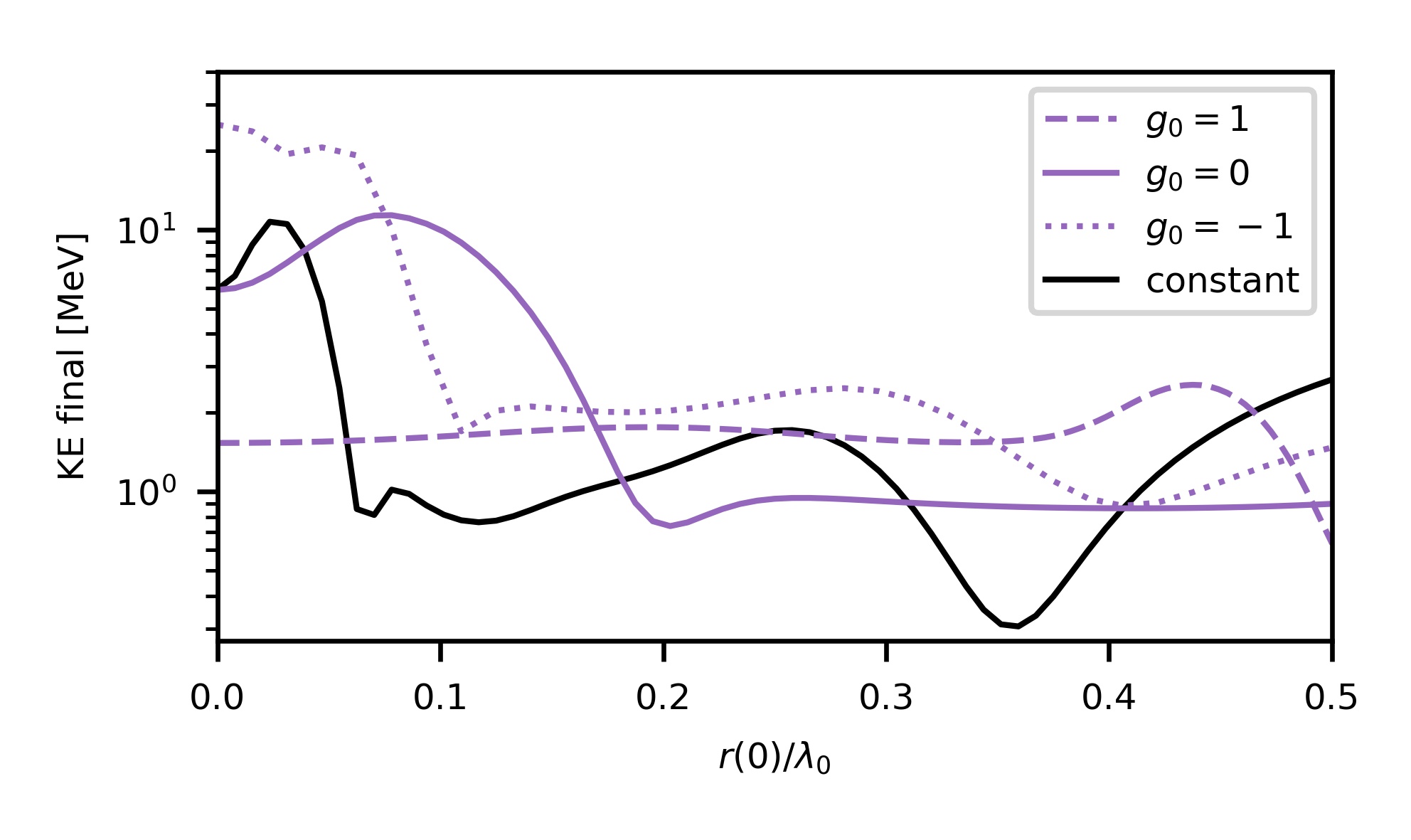}
	\caption{The results of off-axis simulations with varying initial radial position $r(0)$ show a stark difference between cases of varying $g_0$, but also a lack of agreement between $g_0=0$ and the common approximation of constant parameters. Intial electron position $z(0)$ and laser CEP are optimized for the on-axis final energy, and are therefore the same as in Fig.~\ref{fig:fig2}(b) for 125\,TW.}
	\label{fig:2D}
\end{figure}

The model for off-axis acceleration is an extension of the model already used on-axis in Eqs.~(\ref{eq:field})--(\ref{eq:phase}). The off-axis fields including $E_r$ and $B_{\theta}$ must be included along with the non-paraxial terms~\cite{marceau13-1}. We implement this using the fields derived in Ref.~\cite{salamin06} in the frequency domain, which are expanded in terms of the small parameter $\epsilon=w_0/z_R$ that can now depend on frequency. We Fourier-transform the fields to time and calculate the force on the particles as before. We again use a constant beam waist at the central frequency $w_0(\omega_0)=w_{00}=4$\,$\mu$m and vary the $g_0$ value, but to limit the parameter space we simulate a laser pulse with a duration of 3.5\,fs, a power of 125\,TW, and a CEP optimal according to the on-axis accelerated energy (i.e. the same CEP as Fig.~\ref{fig:field}(d)). We simulate the electron trajectories having began at different off-axis positions $r(0)$, but always with the same initial longitudinal position (i.e. the same $z(0)$ as Fig.~\ref{fig:field}(c)). We also compare the properly modelled results with varying $g_0$ to the fully frequency-independent situation (which is not physically correct), where both $w_0$ and $z_R$ are constant values related to the central frequency.

The off-axis simulation results are shown in Fig.~\ref{fig:2D}. It is clear that the results in the different $g_0$ scenarios are also very different off-axis, meaning that the proper modeling is important for the entire process, and not just on-axis trajectories. However, the most important result may be that there is significant difference between the $g_0=0$ case and the results using the frequency-independent approximation. This means that regardless of the $g_0$ value under consideration the frequency-dependent model is necessary to produce correct results for all trajectories. The impact of the differences becomes more signficant especially for applications aiming for electron beams with designed charge or emmitance values, since the effect on the off-axis trajectories is not straightforward.

In conclusion, we have shown that simulating electron acceleration with few-cycle radially-polarized laser fields requires proper modeling of the spatio-spectral amplitude described in this case by the $g_0$ value. This is not only detrimental in the commonly occuring case of $g_0=1$, providing a strong motivation for characterizing experimental driving pulses with spatio-spectral characterization devices~\cite{dorrer19}, but could be a path toward optimizing the acceleration process with shaped amplitude profiles having negative $g_0$, or having a more complex shape that cannot be described by a single parameter. We have shown that these dynamics are due to the varying CEP evolution through the focus in the different cases, and that this need for proper modeling applies to both on-axis and off-axis trajectories. The implications of these results may extend to other areas of very broadband field-sensitive laser-matter interactions such as photoelectron production, high-harmonic generation, and the interaction of intense laser-pulses with solid-density plasmas.

\bigskip

\bibliographystyle{unsrt}

\end{document}